\documentclass[preprintnumbers,showpacs, 
twocolumn,secnumarabic,amssymb,amsmath,amsfonts,aps, pra]{revtex4}

\usepackage{amsmath}
\usepackage{latexsym}
\usepackage{amsfonts}
\usepackage{graphicx}

\begin{document}

\newcommand{\beq}{\begin{equation}}
\newcommand{\eeq}{\end{equation}}
\newcommand{\bea}{\begin{eqnarray}}
\newcommand{\eea}{\end{eqnarray}}
\newcommand{\lps}{\langle}
\newcommand{\rps}{\rangle}
\newcommand{\q}{\mbox{\boldmath $q$}}
\newcommand{\p}{\mbox{\boldmath $p$}}
\newcommand{\Q}{\mbox{\boldmath $Q$}}
\newcommand{\bP}{\mbox{\boldmath $P$}}
\newcommand{\A}{{\bf A}}
\newcommand{\bS}{{\bf S}}
\newcommand{\cn}{\mbox{cn}}
\newcommand{\sn}{\mbox{sn}}


\title{Lyapunov exponents from 
unstable periodic orbits in the FPU-$\beta$ model}

\date{\today}

\author{Roberto Franzosi} 
\email{Roberto.Franzosi@df.unipi.it}
\affiliation{Dipartimento di Fisica Universit\`a di Pisa, and  I.N.F.N., 
Sezione di Pisa, and I.N.F.M., Unit\`a di Pisa, via Buonarroti 2, 
I-56127 Pisa, Italy }

\author{Pietro Poggi}
\email{pietro@dma.unifi.it}
\affiliation{I.F.A.C.-C.N.R., via Panciatichi 64, I-50127 Firenze, Italy}

\author{Monica Cerruti-Sola}
\email{mcs@arcetri.astro.it}
\affiliation{I.N.A.F. - Osservatorio Astrofisico di Arcetri, 
 Largo E. Fermi 5,
 50125 Firenze, and I.N.F.M., Unit\`a di Firenze, 
 Firenze, Italy  }

\begin{abstract}
In the framework of a recently developed theory for Hamiltonian chaos,
which makes use of the formulation of Newtonian dynamics in terms of Riemannian 
differential geometry, we obtained analytic values of 
the largest Lyapunov exponent
for the Fermi-Pasta-Ulam-$\beta$ model (FPU-$\beta$) by computing the time
averages of the metric tensor curvature and of its fluctuations along 
analytically known unstable periodic orbits (UPOs). The agreement between
our results and the Lyapunov exponents obtained by means of standard numerical
simulations supports the fact that UPOs are reliable probes of a general
dynamical property as chaotic instability.  
\end{abstract}

\maketitle

\section{Introduction}

Unstable periodic orbits (UPOs) are widely studied in the framework of 
classical nonlinear dynamical systems \cite{Oono}, since they form
the "skeleton" \cite{Cvitan1} of the phase space of these systems and 
are very sensitive
to local characteristic features of the dynamics.
Among the applications of the studies of UPOs we can mention: 
a characterization of dynamical
systems \cite{Cvitan2}, control of classical chaos, semiclassical quantization
\cite {Cvitan3}. 
Furthermore, they are useful for a characterization of quantum chaos
and for the description of some
thermodynamical properties of dynamical systems with many degrees of freedom. 

The present paper aims at lending further credit to the common wisdom of
relevance of UPOs for chaotic dynamics. Our contribution to this subject
stems from a Riemannian geometric approach to the study of Hamiltonian chaos. 

It is well known that the degree of chaoticity of a dynamical system is 
measured 
by the largest Lyapunov exponent 
$\lambda_1$ which provides an average dynamical instability growth rate 
in terms of the local growth rate of the distance of nearby trajectories,
averaged along a sufficiently long reference trajectory.
The  largest Lyapunov exponent $\lambda_1$ for standard Hamiltonian systems, 
described by Hamiltonian functions of the form 
$H=\sum_{i=1}^N\frac{1}{2}p_i^2 + V(q_1,\dots,q_N)$, 
is computed by numerically integrating the tangent dynamics equation
\begin{equation}
\frac{d^2\xi_i}{dt^2}+ \left(\frac{\partial^2 V}{\partial q^i\partial q^j}
\right)_{q(t)} \xi^j = 0~,
\label{tg-dyn}
\end{equation}
along a reference trajectory
$q(t)=[q_1(t),..,q_N(t)]$, and then $\lambda_1=\lim_{t\to\infty}1/2t \log
 (\Sigma_{i=1}^N[\dot\xi_i^2(t)+\xi_i^2(t)] / \Sigma_{i=1}^N[\dot\xi_i^2(0)+
\xi_i^2(0)] )$.
In the conventional theory of chaos, dynamical instability is caused by
homoclinic intersections of perturbed separatrices, but this theory seems not
adequate to treat chaos in Hamiltonian systems with many degrees of freedom. 
In this case, the direct
numerical simulation is the only way to compute $\lambda_1$.

Recently, it has been proposed by Pettini \cite{Pettini} to tackle Hamiltonian 
chaos in a different theoretical framework with respect to that of homoclinic 
intersections. This new method resorts to a well known formulation of 
Hamiltonian dynamics in the language 
of Riemannian differential geometry: the mechanical trajectories of a dynamical
system can be viewed as geodesics of a Riemannian manifold endowed with a
suitable metric. In this framework, it is possible to relate the instability 
of a geodesics flow with the curvature properties of the underlying 
``mechanical'' manifold through two geometric quantities: the Ricci curvature 
and its fluctuations.
These two geometric quantities, in principle averaged along a generic geodesic,
enter a formula, derived by Pettini et al. in 
Ref.\cite{CasettiClementiPettini},
which allows the analytic computation of the largest Lyapunov exponent for a
generic Hamiltonian system.
However, since the mentioned time averages are in general not analytically 
knowable, one has to replace them with microcanonical averages which  
coincide with time averages when the number of degrees of freedom is large
and the dynamics is chaotic. In fact, under these circumstances, the measure 
of regular orbits in phase space - at physically meaningful energies -
is vanishingly small, thus the dynamics is {\it bona fide} ergodic and 
mixing.

Therefore, the analytic computation of the largest Lyapunov exponent can be done
whenever the simplifying hypotheses of Ref.\cite{CasettiClementiPettini} are
justified and the microcanonical averages of the mentioned geometric
quantities are analytically computable.
This is just the case of the FPU-$\beta$ model which has been considered in Ref.
\cite{CasettiClementiPettini}. 

It is the purpose of the present work to show that, in some special case, the
above mentioned replacement of time averages with microcanonical ones can be
avoided: provided that the {\it time averages} of the Ricci
curvature and of its fluctuations are analytically computed along some unstable 
periodic orbits, a reasonable analytic estimate of the  values of $\lambda_1$
can be obtained without resorting to microcanonical averages. It is somewhat
surprising, and undoubtely very interesting, that unstable periodic orbits make
something like an ``importance sampling'' of the relevant geometric features of 
configuration space which are needed to estimate the average degree of 
chaoticity of the dynamics, measured by $\lambda_1$. 
A similar problem was already addressed in \cite{DRT}, 
where the authors gave an
analytical estimate of the largest Lyapunov exponent at high energy
density for the Fermi-Pasta-Ulam-$\beta$ model by
computing the average of the modulational instability growth rates associated
to unstable modes.

\section{Geometry and dynamics}

Let us summarize the geometrization of Newtonian dynamics tackled in 
\cite{Pettini}.
It applies to standard autonomous systems described 
by the Lagrangian function (all the indices run from $1$ to $N$ degrees of 
freedom)
\begin{equation}
L( q,\dot q)={1\over 2} \sum_{ik} a_{ik}(q)\dot q^i\dot q^k -V( q)~, 
\label{lagr}
\end{equation}
where ~$a_{ik}$~ is the kinetic energy tensor that in terms of 
the total energy $E$ and kinetic energy, reads
\begin{equation}
\sum_{ik} a_{ik}\dot q^i\dot q^k = 2(E - V) = 2 W~,
\label{Ecin}
\end{equation}
Following the method due to Eisenhart \cite{Eisenhart}, the differentiable
$N-$dimensional configuration space ${\cal M}$, on which the lagrangian 
coordinates $(q^1,\ldots,q^N)$ can be used as local coordinates, is enlarged.
The ambient space thus introduced embodies the time coordinate and is
given as ${\cal M} \times {\Bbb R}^2$, with local coordinates
$(q^0, q^1,\ldots,q^N,q^{N+1})$,  where $(q^1,\ldots,q^N)\in {\cal M}$, 
$q^0\in{\Bbb R}$ is the time coordinate, and 
$q^{N+1}\in{\Bbb R}$ is a coordinate closely related to Hamilton 
action.
With Eisenhart we define a pseudo-Riemannian non-degenerate metric 
$g_{{}_E}$ on ${\cal M}\times{\Bbb R}^2$ as
\begin{equation}
\begin{split}
ds_{{}_E}^2=&\sum_{\mu \nu} g_{\mu\nu}\, dq^{\mu} \otimes dq^{\nu} = 
dq^0 \otimes dq^{N+1} + dq^{N+1} \otimes dq^0
+ \\ & 
\sum_{ij} a_{ij} \, dq^i \otimes dq^j 
 -2V(q)\, dq^0 \otimes dq^0~.
\end{split}
\label{g_E}
\end{equation}
Natural motions are now given by the canonical projection $\pi$ of the
geodesics of $({\cal M} \times{\Bbb R}^2, g_E)$ on the configuration 
space-time:
$\pi : {\cal M}\times{\Bbb R}^2\rightarrow {\cal M}\times{\Bbb R}$. 
However, among all the geodesics of $g_{{}_E}$ the natural motions belong 
to the subset of those geodesics along which the arclength is positive 
definite 
\begin{equation}
ds^2 =\sum_{\mu \nu} g_{\mu\nu} dq^\mu dq^\nu = 2C^2 dt^2 > 0 \, ,
\label{par_affine}
\end{equation}
where $C$ is a real arbitrary constant. More details can be found in 
\cite{Pettini}.

The stability of a geodesic flow is studied by means of the
Jacobi$-$Levi-Civita (JLC) equation for geodesic 
spread.
In local coordinates and in terms of proper time $s$ the JLC equation 
reads as
\begin{equation}
\frac{\nabla^2 J^k}{ds^2} + \sum_{ijr}
R^k_{~ijr} \frac{dq^i}{ds}{J^j}\frac{dq^r}{ds} = 0~,
\label{eq_jacobi}
\end{equation}
where $J$ is the Jacobi vector field of geodesic separation, where the 
covariant derivative is given by 
$\nabla J^k/ds= dJ^k/ds + \sum_{ij}\Gamma^k_{ij}\,
dq^i/ds J^j$,
and $R^k_{~ijr}$ are the components of the Riemann-Christoffel curvature 
tensor which, in terms of the Christoffel coefficients $\Gamma^k_{ri}$, are
\begin{equation}
R^k_{~ijr} = \partial_j\Gamma^k_{ri}
-\partial_r\Gamma^k_{ji} +
\sum_{t}
\Gamma^t_{ri}\Gamma^k_{jt}-\Gamma^t_{ji}\Gamma^k_{rt}~
\label{r-ccomp}
\end{equation}
where $\partial_j=\partial/\partial q^j$.
The Christoffel coefficients, in turn, are defined as
\begin{equation}
\Gamma^i_{jk} = \frac{1}{2}\sum_{m} g^{im} \left( \partial_j g_{km} + 
\partial_k g_{mj}
- \partial_m g_{jk} \right)~.
\label{rccoeff}
\end{equation}
In \cite{Pettini} it has been shown that the Jacobi equation (\ref{eq_jacobi}),
written for the Eisenhart metric of the enlarged configuration space, 
nicely yields the standard tangent dynamics equation (\ref{tg-dyn}). Moreover,
under suitable simplifying hypotheses, mainly of geometric type, in 
Ref. \cite{CasettiClementiPettini} it has been shown that Eq.(\ref{eq_jacobi})
can be replaced by a scalar effective equation
\begin{equation}
\frac{d^2 \psi}{ds^2} + \lps k_R\rps_s \psi + \frac{1}{\sqrt{N -1}}
\lps\delta^2 K_R\rps^{1/2}_s \eta (s) \psi = 0~,
\label{eff-eq}
\end{equation}
where $\psi$ stands for any of the components $J^i$ of the Jacobi field,
since in this effective picture all of them obey the same equation.
Moreover, $K_R=\sum_{ijk} g^{ij} R^k_{~ikj}$ is the Ricci curvature and 
$k_R = K_R/ (N -1)$ which, for the Eisenhart metric, 
takes the simple form
\begin{equation}
k_R (q) = \frac{\triangle V}{(N-1)} \simeq \frac{1}{N} \sum_{i=1}^N 
\frac{\partial^2 V(q)}{\partial q^2_{~i}}~.
\label{ricci}
\end{equation}
In Eq.(\ref{eff-eq}), $\eta (s)$ is a gaussian white noise with zero mean and
unit variance, and $\lps \cdot \rps_s$ stands for time averaging along a
reference geodesic. Time averages $\lps k_R\rps_s$ and $\lps\delta^2 K_R\rps_s$
of Ricci curvature and of its second moment respectively, cannot be known
analytically for a chaotic orbit, hence the need of an assumption of ergodicity
allowing the replacement of time averages by
microcanonical averages on a constant energy surface $\Sigma_E$,
corresponding to the energy value $E$ of interest. At variance with time
averages along chaotic orbits, microcanonical averages can be computed
analytically for some models.
It is worth remarking that, after the replacement of time avergaes by means of 
static microcanonical averages $\lps k_R\rps_{\mu_E}$ and 
$\lps\delta^2 K_R\rps_{\mu_E}$, the scalar equation (\ref{eff-eq}) is 
independent of the numerical knowledge of the dynamics.

Then the largest Lyapunov exponent for the effective model given by 
Eq. (\ref{eff-eq}), defined as
\beq
\lambda_1 = \lim_{t \rightarrow \infty} \frac{1}{2t} \log 
\frac{\psi^2(t) + \dot{\psi}^2 (t)}{\psi^2(0) + \dot{\psi}^2 (0)}~,
\label{lyp-eff-eq}
\eeq
is obtained by solving this stochastic differential equation by means of a
standard method due to van Kampen  
\cite{CasettiClementiPettini}, the final analytic expression for $\lambda_1$ 
reads as
\beq
\lambda_1(\Omega_0,\sigma_{\Omega},\tau) =\frac{1}{2} 
\left( \Lambda - \frac{4 \Omega_0}{3 \Lambda} \right)~,
\label{lyp-sol-eff-eq}
\eeq
where $\Omega_0 = \lps k_R \rps_{\mu_E}$, $\sigma^2_\Omega = N \lps \delta^2 
k_R \rps_{\mu_E}$,
\beq
\Lambda = \left( 2 \sigma^2_\Omega \tau + 
\sqrt{ \left( \frac{4 \Omega_0}{3} \right)^3 + 
\left( 2 \sigma^2_\Omega \tau \right)^2}  \right)^{1/3}
\label{Lambd}
\eeq
and 
\beq
2 \tau = \frac{\pi \sqrt{\Omega_0}}
{2 \sqrt{\Omega_0\left( \Omega_0 + \sigma_\Omega \right)} + \pi
\sigma_\Omega}~.
\eeq

\section{Analytic computation of Lyapunov exponents}

In the following, we work out time averages of the Ricci curvature and of
its fluctuations along some analytically known unstable periodic orbits 
of the system described by the Hamiltonian  
\begin{equation}
H(p,q) = \sum_{i= 1}^N \frac{1}{2} p_i^2 +
\sum_{i = 1}^N \left[\frac{1}{2} (q_{i+1} - q_i)^2 + \frac{\beta}{4}
(q_{i+1} - q_i)^4 \right]~,
\label{fpu}
\end{equation}
with periodic boundary conditions $q_{N+1}\equiv q_1$.
This system has been introduced by Fermi, Pasta and Ulam 
in their celebrated work \cite{FPU} on the equipartition properties of the 
dynamics of many non-linearly coupled oscillators. 
Since then, a huge amount
of papers have been devoted to the study of the link beetwen microscopic
 dynamical
properties and macroscopic thermodynamical and statistical properties of 
classical many body systems. 

The linear terms in Hamiltonian (\ref{fpu}) can be diagonalized 
by introducing suitable harmonic normal coordinates. The latters are obtained 
by means of a canonical linear transformation \cite{poggiruffo1}. 
Denoting the normal coordinates and momenta by 
$Q_k$ and $P_k$ for $k = 0,\ldots , N -1$ the transformation is given by
\begin{equation}
Q_k(t) =\sum_{n=1}^N S_{kn} q_k(t)~, \ \
P_k(t) =\sum_{n=1}^N S_{kn} p_k(t)~, 
\label{qtoQ}
\end{equation}
where $k=0,\ldots,N-1$, and  $S_{kn}$ is the ortogonal matrix 
\cite{poggiruffo1} whose elements are
\begin{equation}
S_{kn} = \frac{1}{\sqrt{N}} \left[\sin\left(\frac{2\pi kn}{N}\right) + 
\cos\left(\frac{2\pi kn}{N}\right)\right]~,
\end{equation}
$n =1,\ldots,N$ and $k=0,\ldots,N-1$.
The full Hamiltonian (\ref{fpu}) in the new coordinates reads
\beq
H(\Q,\bP) = \frac{1}{2} P_{0}^2 +
\frac{1}{2} \sum_{i=1}^{N-1} \left(P_{i}^2+\omega_{i}^{2}Q_{i}^2 \right) 
+ H_{1}(\Q)~,
\label{Hfin}
\eeq
where the anharmonic term is
\beq
H_{1}(\Q)= \frac{\beta}{8N} \sum_{i,j,k,l=1}^{N-1}
\omega_{i}\omega_{j}\omega_{k}\omega_{l} C_{ijkl}
 Q_{i}Q_{j}Q_{k}Q_{l}~.
\label{H1fin}
\eeq
The $\omega_{k}=2 \sin ( \pi k/N )$, for $k \in \{1,\ldots,N-1\}$, 
are the normal frequencies for the harmonic case ($\mu=0$), being
$\omega_{k}=\omega_{N-k}$.
By defining
\beq
\Delta_{r}=\left\{ \begin{array}{ll}
(-1)^{m} & \mbox{for $r=mN$ with } m \in {\Bbb Z} \\
0 & \mbox{otherwise~,}
\end{array} \right.
\label{delta}
\eeq
the integer-valued coupling coefficients $C_{ijkl}$ are explicitly given by
\beq
C_{ijkl}=-\Delta_{i+j+k+l}+\Delta_{i+j-k-l}+\Delta_{i-j+k-l}+\Delta_{i-j-k+l}~.
\label{Cijkl}
\eeq
By eliminating the motion of the center of mass (which corresponds to the zero 
index), 
we now easily get the equations of motion for the remaining $N-1$ degrees of 
freedom, which, at the second order, read as
\beq
\ddot{Q}_r =  -\omega_{r}^{2}Q_{r} -
\frac{\beta\omega_{r}}{2N} \sum_{j,k,l=1}^{N-1}
\omega_{j}\omega_{k}\omega_{l}
C_{rjkl} Q_{j}Q_{k}Q_{l}~,
\label{eqmoto}
\eeq
for $r=1,\ldots,N-1$.

As is shown in Ref.\cite{poggiruffo1}, the equations of motion (\ref{eqmoto}) 
admit some exact, periodic solutions that can be explicitly 
expressed in closed analytical form.
The simplest ones, consisting of one mode (OM), have only one
excited mode, which we denote by the index $e$, and thus
are characterized by $Q_j(t)\equiv 0$ for $j\neq e$. 
The solitary modes are found by setting
$C_{reee} = 0 \ \forall r \in  \{1,\ldots,N-1\}$ with $r \neq e$; 
it is easily verified that this condition is satisfied for 
\beq
e=\frac{N}{4};\: \frac{N}{3};\: \frac{N}{2};
\: \frac{2N}{3};\: \frac{3N}{4}~.
\label{solitari}
\eeq
Thus, for solutions with initial conditions $Q_j=0$ and $\dot{Q}_j=0$
for $j \neq e$, the whole system (\ref{eqmoto})
reduces to a one degree of freedom (and thus integrable) system described by
the equation of motion 
\beq
\ddot{Q}_{e} = - \omega_e^2 Q_{e} -
\frac{\beta \omega_{e}^4 C_{eeee}}{2 N} Q_{e}^3~,
\label{eqe}
\eeq
where $C_{eeee}= 4, 4, 3, 3, 2$ for $e=N/4, 3N/4, N/3, 
2N/3$, $N/2$, respectively.
The harmonic frequencies of the modes~(\ref{solitari}) are 
$\omega_e =\sqrt{2},\sqrt{2},\sqrt{3},\sqrt{3},2$ for 
$e=N/4,$ $3N/4,N/3,2N/3,N/2$, respectively.
In order to simplify the notation, in the following, let us set 
$\hat{C}_e = C_{eeee}$.

The general solution of (\ref{eqe}) is a Jacobi elliptic cosine
\beq
Q_{e} (t) = A \: \cn \left[ \Omega_e (t -t_0) \, ,\, k  \right]~,
\label{e}
\eeq
where the free parameters (modal) amplitude $A$ and time origin $t_0$
are fixed by the initial conditions. The frequency $\Omega_e$ and the 
modulus $k$ of Jacobi elliptic cosine function \footnote{For standard notation 
and properties of elliptic functions and integrals we refer 
to: D. F. Lawden {\it Elliptic Functions and Applications},
Springer-Verlag, New York (1989) and
{\it Handbook of Mathematical Functions},
M. Abramowitz and I. A. Stegun
eds., Dover, New York (1965).}
depend on $A$ as follows
\beq
\Omega_e =  \omega_e  \sqrt{1+\delta_e  A^2} \ ,\quad
k = \sqrt{\frac{\delta_e A^2}{2 (1 +\delta_e A^2)}}~~, 
\label{modulus}
\eeq
with $\delta_e=\beta \omega_e^2 \hat{C}_e/(2N)$.
This kind of solution is periodic, and its oscillation period $T_e$
depends on the amplitude $A$, since it is given in terms of
the complete elliptic integral of the first kind ${\bf K}(k)$ and 
in terms of $\Omega_e$ by
\beq
T_e = \frac{4 {\bf K}(k)}{\Omega_e}~.
\label{period}
\eeq
The modal amplitude $A$ is one-to-one related to the energy density 
$\epsilon = E/N$. In fact, computing the total energy~(\ref{Hfin})
on the OM solution $Q_j(t)\equiv\delta_{je} Q_e(t)$, one finds
\beq
\epsilon N = \frac{1}{2} \left( P_e^2 + \omega_e^2 Q_e^2 \right) +
\frac{\beta}{8N} \omega_e^4 \hat{C}_e Q_e^4~.
\eeq
Since at $t=t_0$ the coordinates result $(Q_e(t_0),P_e(t_0))=(A,0)$, 
by solving the previous equation for $A$ we get
\beq
A= \left[ 2N \left(\frac{\sqrt{1+ 2 \beta \epsilon \hat{C}_e} - 1}
{\beta \omega_e^2 \hat{C}_e} \right) \right]^{1/2}~.
\label{Afrome}
\eeq
This relation allows to express all the parameters of the solution~(\ref{e})
in terms of the more physically relevant parameter $\epsilon$.
The period $T_e$ is
\beq
T_e = \frac{4 {\bf K}(k)}{\omega_e (1 + 2 \beta \epsilon \hat{C}_e)^{1/4}}~,
\label{period2}
\eeq
where $k=k(\epsilon)$ can be found from~(\ref{modulus}) and~(\ref{Afrome}).

In terms of the standard coordinates, the OM solutions result
\beq
q_n(t) = \frac{1}{\sqrt{N}} Q_e(t) \left[ \sin \left( \frac{2\pi n e}{N}
\right) + \cos \left( \frac{2\pi n e}{N} \right) \right]~,
\label{qtoQ2}
\eeq
where $e$ is one of the values listed in~(\ref{solitari}).

\begin{figure}[ht]
\includegraphics[scale=0.45]{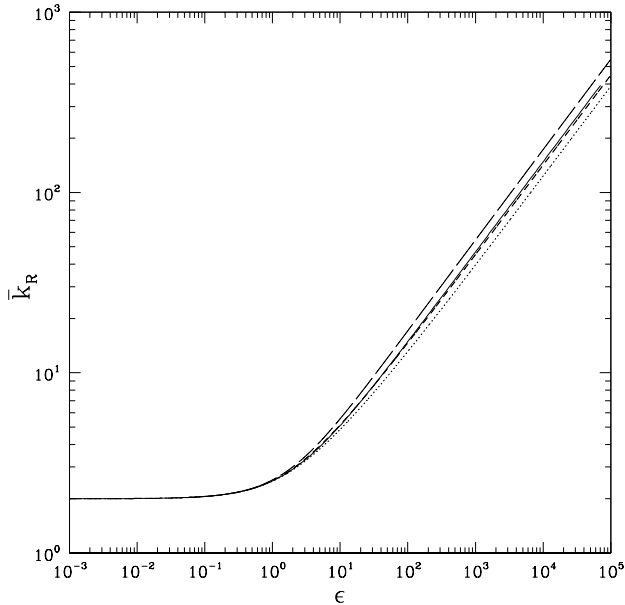}
\caption{$\overline{k}_R$ versus $\epsilon$, worked out by means of
the three single mode solutions identified by the values of 
$e$ listed in~(\ref{solitari})
(dotted, dashed and
long-dashed lines refer to $e=N/4,3N/4$, $e=N/3,2N/3$ and $e=N/2$,
respectively),
 is compared with $\lps k_R\rps_{\mu_E}$
computed in 
\cite{CasettiClementiPettini} (continuous line).
 The agreement is very good on a broad range of
values of energy density $\epsilon$.}
\label{kr}
\end{figure}

The Ricci curvature along a periodic trajectory, obtained by substituting
Eq. (\ref{qtoQ2}) into Eq. (\ref{ricci}), is 
\beq
k_R(t) = 2 + \frac{6 \beta}{N} \omega^2_{e} Q^2_{e}(t)~,
\eeq
and we can compute its time average $\overline{k}_R$ as 
\beq
 \overline{k}_R= 2 +\frac{6 \beta}{N} \omega^2_{e} \overline{Q^2}_{e}~.
\label{krfromQ}
\eeq
After simple algebra, using standard properties of the elliptic functions, we 
find
\beq
\overline{Q^2}_{e} = \frac{1}{T_e} \int_{t_0}^{T_e+t_0} dt\ Q^2_{e} = 
\frac{A^2}{{\bf K} k^2} \left( {\bf E} + (k^2 -1) {\bf K} \right)~.
\label{Q_2_end} 
\eeq
The time averaged Ricci curvature results
\beq
\overline{k}_R = 2 + \frac{12}{{\bf K} k^2 \hat{C}_e} 
\left[ \sqrt{1 + 2 \beta \epsilon \hat{C}_e} - 1 \right]\! 
\left[ {\bf E} + (k^2 - 1) {\bf K} \right]~,
\label{krlast}
\eeq
where {\bf K} and {\bf E} are the complete elliptic integrals of the first
and second kind respectively, both depending on the modulus $k$ which,
from~(\ref{modulus}) and~(\ref{Afrome}), is determined by the 
energy density $\epsilon$ 
\beq
k^2 = \frac{1}{2} \left( 1 - \frac{1}{\sqrt{1 + 2 \beta \epsilon \hat{C}_e}}
\right)
\label{modulusfrome}~.
\eeq
Now, using Eqs. (\ref{krlast}) and (\ref{modulusfrome}), and the tabulated 
values for ${\bf E}$ and ${\bf K}$, $\overline{k}_R$ is given as a function
of the energy density $\epsilon$.
In Fig. \ref{kr} a comparison is made between $\overline{k}_R$ versus 
$\epsilon$, worked out for
the OM solutions under consideration, and $\langle k_R\rangle_{\mu_E}$ versus 
$\epsilon$, the average Ricci curvature analytically computed in Ref.
\cite{CasettiClementiPettini}.

By definition, the average of the curvature fluctuations is 
\beq
\begin{split}
\lps\delta^2 K_R\rps_\mu =& \left< (K_R - \lps K_R\rps_\mu)^2 \right>_\mu = \\
&
(N-1)^2 \left[ \lps(k_R)^2\rps_\mu - (\lps k_R\rps_\mu)^2 \right]~.
\end{split}
\eeq
Again, by replacing the microcanonical averages with time averages,
from Eq. (\ref{krfromQ}) and after some trivial algebra, we get
\beq
\overline{\delta^2 k_{R}} = 
\frac{36 \beta^2 \omega^4_{e}}{N^2} \left[ \overline{Q^4}_{e} -  
\overline{Q^2}_{e}~\overline{Q^2}_{e}  \right]~.
\label{dk2r}
\eeq
The new term
\[
\overline{Q_{e}^4} = \frac{A^4}{T_e} \int_0^{T_e} dt\ \cn^4(\Omega_e t, k) = 
\frac{A^4}{4 {\bf K}} \int_0^{4 {\bf K}} d\theta\ \cn^4 (\theta, k) 
\]
can be computed by resorting to standard properties of the elliptic functions 
and the result is
\beq
\overline{Q_{e}^4} =\frac{A^4}{3 {\bf K} k^4} \left[ {\bf K} 
( 2 - 5 k^2 + 3 k^4 )
+ 2 {\bf E} (2 k^2 - 1) \right]~.
\label{Q4}
\eeq
Finally, Eqs. (\ref{Q4}) and (\ref{Q_2_end}) in (\ref{dk2r}) yield
\beq
\overline{\delta^2 k_{R}} = \frac{192\left[ (k^2 - 1) + 2(2 - k^2)\frac{{\bf E}}{{\bf K}}  - 
3 \left( \frac{{\bf E}}{{\bf K}} \right)^2 \right]}{(1-2 k^2)^2 \hat{C}^2_{e}}
 ~.
\label{dk2re}
\eeq
From Eq. (\ref{modulusfrome}) and making use of the tabulated values
for ${\bf E}$ and ${\bf K}$, equation~(\ref{dk2re}) provides the 
mean fluctuations of curvature as a function of $\epsilon$.

\begin{figure}[ht]
\includegraphics[scale=0.45]{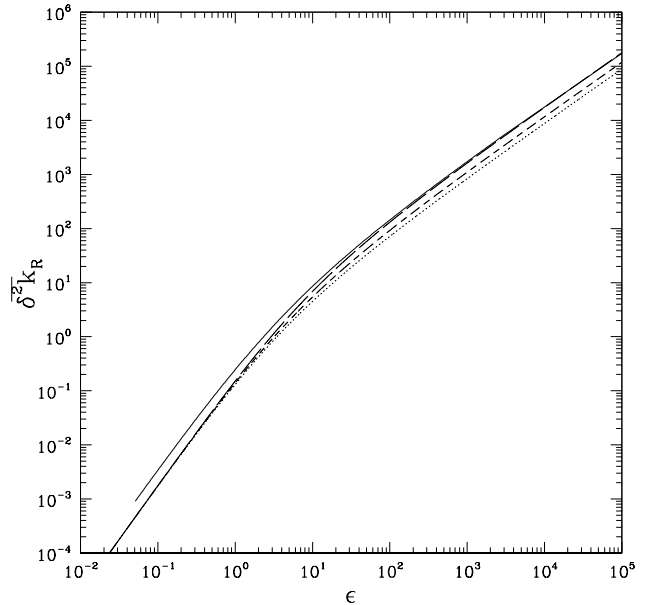}
\caption{In this figure we report three curves for $\overline{\delta^2 k}_R$
versus $\epsilon$ computed by integrating the curvature fluctuations
along the three single mode solutions considered in the present paper
(dotted, dashed and
long-dashed lines refer to $e=N/4,3N/4$, $e=N/3,2N/3$ and $e=N/2$,
respectively),
and a comparison is made with
the same quantity computed in \cite{CasettiClementiPettini}(continuous line).
Also in this case the agreement is very good.}
\label{d2kr}
\end{figure}
In Fig. \ref{d2kr}, a comparison is made between the time average of the Ricci
curvature fluctuations $\overline{\delta^2 k_{R}}$ as a function of the energy 
density $\epsilon$, 
worked out along the OM solution that we considered, and 
$\langle\delta^2 k_R\rangle_{\mu_E}$ versus $\epsilon$
analytically computed in Ref. \cite{CasettiClementiPettini}. 
The agreement is very good, thus confirming from a completely new point 
of view, that 
unstable periodic orbits are special tools for dynamical systems analysis; in
this case, certain geometric quantities of configuration space are surprisingly 
well sampled by UPOs because time averages computed along them are very close
to microcanonical averages performed on the whole energy hypersurfaces. 

\begin{figure}[t]
\includegraphics[scale=0.45]{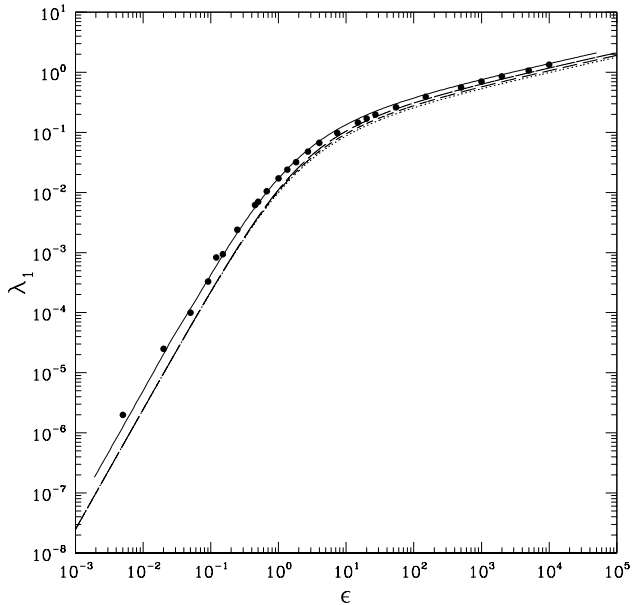}
\caption{This figure shows the largest Lyapunov exponent $\lambda_1$
obtained by integrating the suitable
geometric quantity along the three single mode
solutions considered in the present paper, plotted vs. $\epsilon$.
Dotted, dashed and
long-dashed lines refer to $e=N/4,3N/4$, $e=N/3,2N/3$ and $e=N/2$,
respectively.
Continuous line refers to the Lyapunov exponent computed
in \cite{CasettiClementiPettini}.  The full circles are the values
for $\lambda_1$ computed by
numerical integration. The agreement is  again very good on a broad range
of $\epsilon$ values.}
\label{lyap}
\end{figure}

Finally, we can compute the Lyapunov exponents as a function of the energy
density $\epsilon$ by inserting Eqs. (\ref{krlast}), (\ref{modulusfrome}) and
(\ref{dk2re}) into the analytic formulae \ref{lyp-sol-eff-eq} and \ref{Lambd},
replacing $\langle k_R\rangle_{\mu_E}$ and $\langle\delta^2 k_R\rangle_{\mu_E}$
by means of the corresponding time averages computed above.
Fig. \ref{lyap} shows that the overall agreement between our analytic results,
the analytic results from \cite{CasettiClementiPettini} and the results
obtained by numerical integration of the tangent dynamics, is very good.
The agreement is globally very good because at high energy density our results
are really very close to the other mentioned ones, and at low energy density 
the discrepancy does not exceed -- at worst -- a factor of $2$ on a range of
many decades of energy density and with the use of only {\it one} unstable 
periodic orbit! 

\section{Concluding remarks}

In conclusion, we have found that some global curvature properties of the
configuration space manifold -- whose geodesics coincide with the trajectories
of an Hamiltonian system -- are well sampled by unstable 
periodic orbits. Then, since the averages of these curvature quantities enter 
an analytic formula to compute the largest Lyapunov exponent, unstable periodic 
orbits can be used also to compute Lyapunov exponents through the time averages 
of the same geometric quantities.
In the present work, this result has been obtained in the case of the 
FPU-$\beta$ model for which the analytic expression of some unstable periodic 
solutions of the equations of motion are known. The outcome of this computations
is in very good agreement with those reported in
Ref.\cite{CasettiClementiPettini} on $\lambda_1$ for the same model. 
Of course, it would be very interesting to perform similar computations also
for other models.
Finally, from a new point of view, that of the Riemannian geometric theory of
Hamitonian chaos, we confirm that unstable periodic orbits seem to have a 
special relevance among all the possible phase space trajectories of a 
nonlinear Hamiltonian system.


\begin{references}

\bibitem{Oono} Y. Oono, Prog. Theor. Phys. {\bf 59}, 1028 (1978).

\bibitem{Cvitan1} P. Cvitanovi\v c, Physica D {\bf 51}, 138 (1991).

\bibitem{Cvitan2} P. Cvitanovi\v c, B. Eckhardt, J. Phys. A {\bf 24}, 
L237 (1991).

\bibitem{Cvitan3} P. Cvitanovi\v c, B. Eckhardt, Phys, Rev. Lett. 
{\bf 63}, 823
(1989).

\bibitem{Pettini} M. Pettini, Phys. Rev. E {\bf 47}, 828 (1993). For a recent
review, see L. Casetti, M. Pettini and E.G.D. Cohen, Phys. Rep.
{\bf 337}, 237 (2000), and references therein quoted.

\bibitem{CasettiClementiPettini} L.Casetti, C. Clementi and M. Pettini,
Phys. Rev. E {\bf 54}, 5969 (1996).

\bibitem{DRT} T. Dauxois, S. Ruffo and A. Torcini, Phys. Rev. E
{\bf 56}, R6229 (1997).

\bibitem{Eisenhart} L. P. Eisenhart, Ann. Math. {\bf 30}, 591 (1929).

\bibitem{FPU} E. Fermi, J. Pasta, and S. Ulam, 
	Los Alamos Report LA-1940 (1955), in {\it Collected papers of Enrico 
	Fermi}, edited by E. Segr\'e, (University of Chicago, Chicago,
	1965), Vol. 2, p. 978.

\bibitem{poggiruffo1} P. Poggi and S. Ruffo, Physica D {\bf 103}, 251 (1997).



\end{references}
\end{document}